\documentclass[entropy,article,accept,moreauthors,pdfm,10pt,a4paper]{mdpi} 

\firstpage{1} 
\doinum{10.3390/e23010055}
\articlenumber{1}
\pubvolume{23}
\pubyear{2021}
\history{Received: 13 December 2020; Accepted: 29 December 2020; Published: 31 December 2020}
\firstpage{1} 
\lastpage{\pageref*{LastPage}}
\doinum{10.3390/e23010055}
\articlenumber{1}
\pubvolume{23}
\pubyear{2021}
\history{Received: 13 December 2020; Accepted: 29 December 2020; Published: 31 December 2020}

\Title{Applicability of squeezed- and coherent-state continuous-variable quantum key distribution over satellite links}

\Author{Ivan Derkach $^{\dagger}$ and Vladyslav C. Usenko $^{\dagger}$}

\address[1]{%
Department of Optics, Palacky University, 17. listopadu 12, 771 46 Olomouc, Czech Republic}

\corres{Correspondence: usenko@optics.upol.cz;}

\firstnote{These authors contributed equally to this work.}

\abstract{We address the applicability of quantum key distribution with continuous-variable coherent and squeezed states over long-distance satellite-based links, considering low Earth orbits and taking into account strong varying channel attenuation, atmospheric turbulence and finite data ensemble size effects. We obtain tight security bounds on the untrusted excess noise on the channel output, which suggest that substantial efforts aimed at setup stabilization and reduction of noise and loss are required, or the protocols can be realistically implemented over satellite links once either individual or passive collective attacks are assumed. Furthermore, splitting the satellite pass into discrete segments and extracting the key from each rather than from the overall single pass allows to effectively improve robustness against the untrusted channel noise and establish a secure key under active collective attacks.  We show that feasible amounts of optimized signal squeezing can substantially improve the applicability of the protocols allowing for lower system clock rates and aperture sizes and resulting in higher robustness against channel attenuation and noise compared to the coherent-state protocol.}


\begin{document}


    \section{Introduction}
Quantum key distribution (QKD) \cite{Diamanti2016,xu2020secure, pirandolaadvances} is well known to have its goal in developing methods (protocols) for sharing a secret key between legitimate users, who can lately use the key for the confidential information transfers. First started with the discrete-variable protocols based on direct detection of single-photon states (and their emulation using weak coherent pulses or entangled photon pairs \cite{Gisin2002}), QKD was later extended to the realm of continuous variables (CV) \cite{Braunstein2005} based on efficient and low-noise homodyne detection of multiphoton coherent or squeezed states of light. 

One of the important applications of QKD is in the extra-terrestrial channels, which potentially allow extremely long-distance secure communication enabled by QKD over a satellite. While discrete-variable protocols were recently successfully tested over the satellite links \cite{vallone2015experimental, Liao2018, villar2020entanglement, yin2020entanglement}, the applicability of satellite-based CV QKD remains less studied. Indeed, it was considered in the asymptotic regime of the infinitely many quantum states \cite{Hosseinidehaj2016, Hosseinidehaj2019}, which however is never the case in practice. Moreover, CV QKD may have an important practical advantage in free-space applications and particularly in the satellite-based channels because a homodyne detector, in which the signal is coupled to a narrow-band local oscillator (bright coherent beam used as a phase reference), intrinsically filters out the background radiation at unmatched wavelengths \cite{Elser2009}. Thus CV QKD can operate in conditions of strong stray light and potentially at daytime, which, in the case of discrete-variable protocols, would require additional filtering, increasing attenuation and complexity of the set-up. Recently, the feasibility of coherent-state CV QKD over satellite links was discussed in \cite{dequal2020feasibility}. In the current work we analyze applicability of CV QKD over satellite-based channels considering also squeezed signal states. As the feasible squeezing up 10 dB, achievable with current technology \cite{andersen201630}, is known to improve robustness of CV QKD to noise \cite{garcia2009continuous, Madsen2012, Usenko2011}, we confirm its usefulness in the satellite-based links as well. We build the channel model on the assumption of normal fluctuation of deflected signal beam center around the receiving aperture center, and study applicability of CV QKD, taking into account the finite data ensemble size. We show that in this regime the protocols appear to be extremely sensitive to strong channel attenuation and large amounts of data are required for successful realization of CV QKD over satellites, which contradicts relatively short passage times. Possible solutions to circumvent the problem can be i) use of squeezed states, that reduce requirements on the data ensemble size and can tolerate stronger attenuation and channel noise; ii) relaxation of security assumptions considering individual attacks or passive eavesdropping, introducing no excess noise; iii) increase of the link transmittance using larger telescopes in the downlink regime; iv) increase of the repetition rate of the system in order to accumulate larger statistics. Our results reveal substantial challenges for implementing CV QKD over satellites, but shows no fundamental limits for such realizations. Already under the strict assumption of collective eavesdropping attacks and untrusted channel noise, CV QKD protocols using feasible squeezing should be applicable with low-orbit satellites, while in the assumption of passive eavesdropping squeezed-state CV QKD can tolerate up to 43 dB of channel attenuation, which paves the way towards realization over geostationary satellites.


    \section{Security of CV QKD}
\begin{figure}
        \centering
        \includegraphics[width=.5\linewidth]{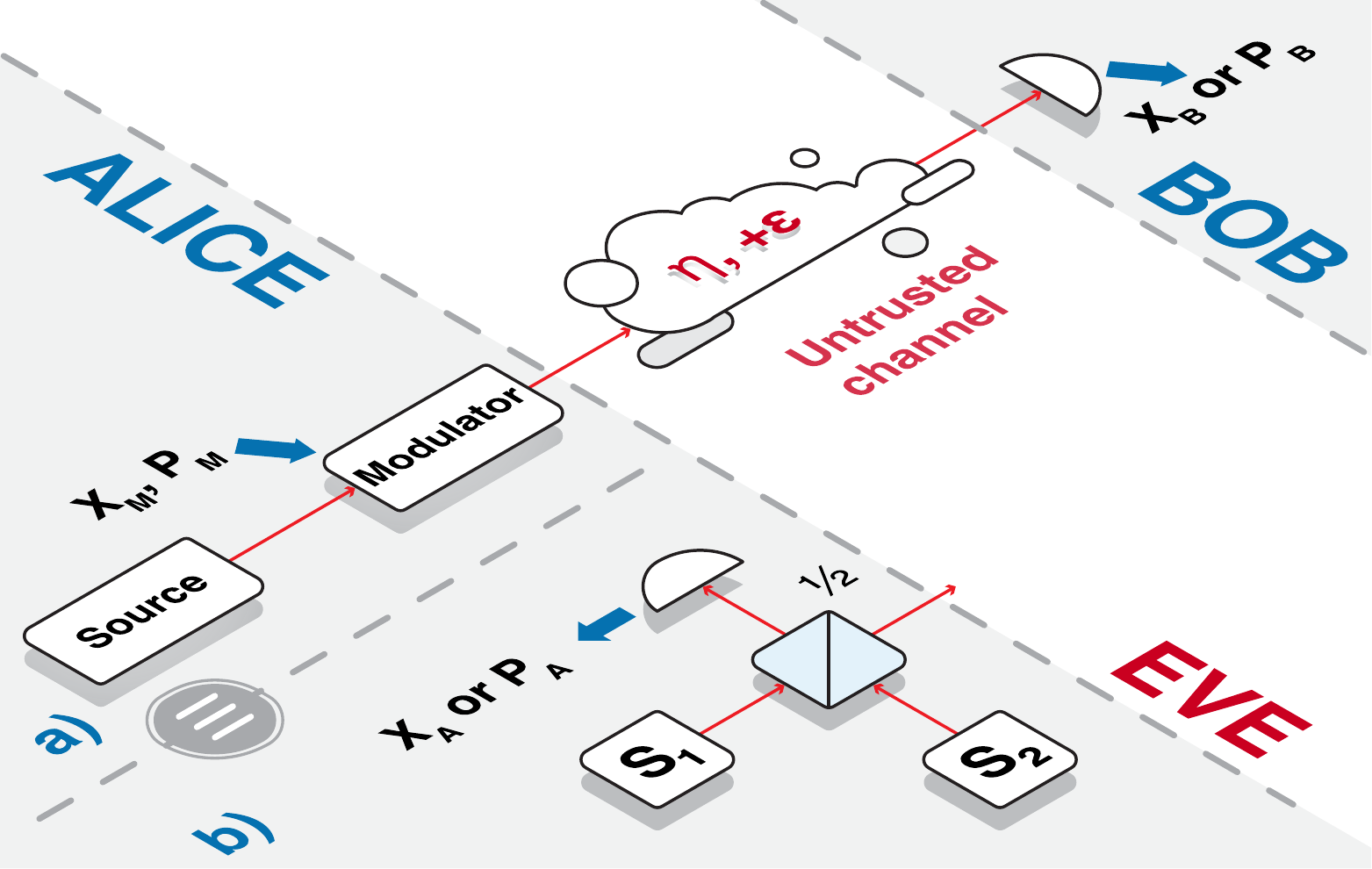}
\caption{a) CV QKD scheme based on a signal state preparation using a Source (laser or optical parametric oscillator for preparation of coherent or squeezed states respectively) and a quadrature Modulator (driven by the data $x_M,p_M$) on the side of Alice and on a homodyne detection on the side of Bob, resulting in measurement outcomes $x_B$ or $p_B$ after the propagation through an untrusted quantum channel with transmittance $\eta$ and excess noise $\epsilon$ related to the channel output; b) Theoretical purification scheme of the state preparation on the side of Alice based on generation of two oppositely squeezed states with variances (\ref{varexp}) on optical parametric oscillators (squeezers) $S_1$ and $S_2$, coupling squeezed states on a balanced beamsplitter, and local homodyne measurement on the Alice's side resulting in $x_A$ or $p_A$ on one of the modes (equivalent to modulation of squeezed states; in the case of coherent-state protocol, a heterodyne detection, resulting in $x_A$ and $p_A$, is considered at Alice's side), while the other mode is sent to the channel towards Bob, the rest of the scheme is an in a).
        \label{scheme}}
    \end{figure}

We address satellite-based implementation of Gaussian CV QKD protocols \cite{Weedbrook2012} using coherent or squeezed states of light as shown in Figure \ref{scheme} (a). We describe the quantum states of light in a given mode of electromagnetic radiation using two complementary observables, namely quadratures, being analogues of position and momentum operators of a single particle, and expressed through mode's quantum operators as $\hat{x}=\hat{a}^\dag+\hat{a}$ and $\hat{p}=i[\hat{a}^\dag-\hat{a}]$. The sender Alice prepares coherent or squeezed states using respectively a laser source or an optical parametric oscillator \cite{Giordmaine1965}, and modulates the states by applying quadrature displacements, governed by independent zero-centered Gaussian distributions, by using quadrature modulators. This way Alice prepares the states described by the quadratures $\hat{x}_A=\hat{x}_S+\hat{x}_M$ and $\hat{p}_A=\hat{p}_S+\hat{p}_M$, where $\hat{x}_S$ and $\hat{p}_S$ with variances $Var(\hat{x}_S)=V_S$ and $Var(\hat{p}_S)=1/V_S$ (we define variance of an operator $\hat{r}$ with zero mean value as $Var(\hat{r})=\langle \hat{r}^2 \rangle$) are the quadrature values of the signal (so that either $V_S=1$ for coherent states or with no loss of generality we assume $x-$quadrature squeezed states with $V_S<1$). $\hat{x}_M$ and $\hat{p}_M$ with $Var(\hat{x}_M)=Var(\hat{p}_M)=V_M$ are the displacements known to Alice, which constitute her classical data contributing to the final secret key. The signal then travels through a generally noisy and lossy quantum channel, which can be optimally \cite{Navascues2006,Garcia2006} represented as a Gaussian channel resulting in the output state described by the quadratures $\hat{x}_B=\sqrt{\eta}\hat{x}_A+\sqrt{1-\eta}\hat{x}_0+\hat{x}_N$, where $\eta$ is the channel transmittance, $\hat{x}_0$ with $Var(\hat{x}_0)=1$ is the variance of the vacuum noise concerned with the channel attenuation, and $\hat{x}_N$ is the contribution from the excess noise on the channel output with $Var(\hat{x}_N)=\epsilon$, similarly for $p-$quadrature with the same $\eta$ and $\epsilon$, as the free-space channel is typically phase-insensitive. Trusted parties use a beacon laser for signal acquiring, tracing and pointing, as well as for timing synchronization \cite{liao2017satellite}. The receiving side Bob is performing homodyne detection on the incoming mode by measuring $\hat{x}_B$ or $\hat{p}_B$ (bases have to be randomly switched between in order to fully characterize channel loss and excess noise). 
Preferably such measurement is made using locally generated local oscillator \cite{soh2015self, qi2015generating}, that despite additional noise contribution from signal wavefront aberration \cite{kish2020use} and phase noise due to relative phase drift \cite{marie2017self, kleis2017continuous, laudenbach2019pilot}, allows to avoid phase reference pulse attenuation in a strongly lossy satellite link, and provides enhanced security of the protocol by ruling out the attacks on the local oscillator.
After accumulating a certain amount of data points $N$ from state preparation and measurement, parties use $N-n$ of these to estimate the channel parameters $\eta$ and $\epsilon$, from which the security of the protocol can be assessed as described below. The remaining $n$ points are processed using error correction and privacy amplification algorithms \cite{Gisin2002} in order to obtain the resulting provably secure key which can then be used for classical encryption. The ratio $n/N$ can be optimized \cite{Ruppert2014}, here for simplicity we assume it to be $1/2$ (which is close to optimal except for the low repetition rates) and resulting estimates to be perfectly accurate \textit{i.e.} with infinitesimal confidence intervals (or already being pessimistic lower bounds complying with a given probability of failure of the channel estimation procedure). We assume that the remote trusted side (Bob) is the reference side for the error correction algorithms, thus using so-called reverse reconciliation, which was shown robust against any level of pure channel loss \cite{Grosshans2003} and is applicable in the strongly attenuating links with $\eta$$\ll 1$.
 
 Security of CV QKD was established against two main types of attacks, namely individual, when an eavesdropper is able to interact optimal probe states with the signal states and then individually measure the probes, and collective, when the probes are assumed to be stored in a quantum memory after the interaction and then optimally collectively measured, which increases the amount of information, that can be potentially obtained by an eavesdropper \cite{Diamanti2015}. 
Security of the coherent-state protocol with heterodyne detection against collective can be extended to security against general attacks using de Finetti reduction \cite{leverrier2017security}, similar extensions for the squeezed-state protocol and homodyne detection being more demanding  \cite{furrer2012continuous, leverrier2013security}.

We study security of the above described protocol by evaluating the lower bound on secure key rate per channel use, which in the finite-size regime, reverse reconciliation scenario, and firstly assuming collective attacks, performed by an eavesdropper, reads

\begin{equation}
K=\text{max}\left\{0,\frac{n}{N}\Big[\beta I_{AB}-\chi_{BE}-\delta(n)\Big]\right\},
\label{kr}
\end{equation}

where $\beta \in [0,1]$ is post-processing efficiency representing how close the trusted parties are able to reach the Shannon mutual information $I_{AB}$ using realistic error correction codes, $\chi_{BE}$ is the Holevo bound \cite{holevo2001evaluating}, giving the upper bound on an eavesdropper's information on the measurement results at the remote receiving side, Bob, and $\delta(n) \approx 7\sqrt{\frac{\log_2{(2/\bar{\epsilon})}}{n}}$ is the correction parameter related to the finite-size effects \cite{Leverrier2010}, where $\bar{\epsilon}$ is the 
smoothing parameter that contributes to the overall failure probability of the protocol, and is set further to $10^{-10}$. 

The mutual information between the trusted parties $I_{AB}=1/2\log_2{V_B/V_{B|A}}$ for the Gaussian-distributed data can be expressed through variances and correlations between the measurement outcomes, \textit{i.e.}, through the variance of Bob's measurement $V_B=\eta V+(1-\eta)+\epsilon$, where $V\equiv V_S+V_M$, and conditional variance $V_{B|A}=V_B-C_{AB}^2/V_A$, $V_A \equiv V_M$ is the variance of Alice's data and $C_{AB}=\sqrt{\eta}V_M$ is the correlation between modulation data and measurements on $\hat{x_B}$, for the zero-mean-distributed observables obtained as $C_{AB}=\langle \hat{x}_M\hat{x}_B \rangle$. The resulting expression for the mutual information then reads

\begin{equation}
I_{AB}=\frac{1}{2}\log_2{\Big[1+\frac{\eta V_M}{\eta V_S+1-\eta+\epsilon}\Big]}
\label{eq:iab}
\end{equation}

and is essentially determined by the signal modulation variance $V_M$, signal state variance $V_S$, channel transmittance $\eta$, and channel noise $\epsilon$ related to the channel output. 

Evaluation of the Holevo bound on the other hand is more involved. It is based on the assumption that Eve holds purification of the noise added in the channel (due to losses and excess noise) and relies on the evaluation of von Neumann entropies of the state shared between Alice and Bob \cite{Usenko2016}. This is performed in the equivalent entanglement-based representation, when state preparation is purified using two-mode entangled state. In the case of coherent-state protocol a symmetrical two-mode squeezed vacuum (TMSV) \cite{Ou1992} state with variance $V=1+V_M$ is used for purification, such that Alice is measuring one of the modes using a heterodyne (balanced homodyne) detector \cite{Usenko2016}. For a general state preparation in the squeezed-state protocol, assuming independent levels of signal squeezing and modulation variance (contrary to the standard symmetrically modulated protocol, where squeezing and modulation are essentially related as $V_M=1/V_S-V_S$ \cite{Cerf2001}), we use the generalized entanglement-based scheme using an asymmetrical entangled state instead of TMSV and a homodyne detection on the local mode \cite{Usenko2011}, as shown in Figure \ref{scheme} (b). The preparation in this case is equivalent to the prepare-and-measure scheme provided the asymmetrical state is constructed of the oppositely squeezed states with variances in $x$-quadratures being

    \begin{eqnarray}
        V_1=V_S+V_M-\sqrt{V_M(V_S+V_M)},\hspace{1cm}
        V_2=1/\big[V_S+V_M+\sqrt{V_M(V_S+V_M)}\big],
    \label{varexp}
    \end{eqnarray}

while having the opposite variances ($1/V_1$ and $1/V_2$ respectively) in the $p$-quadratures.

In the purification-based scenario, the Holevo bound is evaluated as $\chi_{BE}=S(AB)-S(A|B)$, where $S(\cdot)$ denotes the von Neumann (quantum) entropy of a state, $S(AB)$ is the quantum entropy of a (generally noisy) state shared between the trusted parties and $S(A|B)=S(A|x_B)$ is the von Neumann entropy of the state of the trusted parties, conditioned by Bob's measurement results in $x$-quadrature. We obtain the relevant von Neumann entropies from bosonic entropic functions of symplectic eigenvalues of respective covariance matrices of the states, shared between Alice and Bob (see \cite{Usenko2016} for details of security analysis techniques in CV QKD).

In the case of individual attacks and reverse reconciliation scenario, the upper bound on the information leakage is reduced to the classical (Shannon) information between Bob and Eve, which, similarly to $I_{AB}$ (\ref{eq:iab}) described above, reads $I_{BE}=(1/2)\log_2{(V_B/V_{B|E})}$. The evaluation of $V_{B|E}$ is done in the assumption that Eve is able to purify the channel noise. The optimal individual attack in this case is the entangling cloner attack \cite{Grosshans2003a}, which is a TMSV state of variance $V_N=1+\frac{\epsilon}{1-\eta}$ set so to emulate the channel loss $\eta$ and noise $\epsilon$. One of the modes of the cloner interacts with the signal with a linear coupling $\eta$, corresponding to the channel loss, and resulting in $V_B=\eta V+(1-\eta)V_N$ (which gives exactly $V_B$ as described above and as expected by the trusted parties), while the other mode is measured by Eve in order to reduce her uncertainty on the quantum noise added in the channel. The mutual information between Bob and Eve then reads \cite{Grosshans2003a, garcia2007quantum}

\begin{equation}
I_{BE}=\frac{1}{2}\log_2\left\{\frac{1}{V}[1+\epsilon+\eta(V-1)][\eta(V-1)+V(\epsilon+1)]\right\},
\end{equation}

which gives the bounds on the secure key rate in the case of individual eavesdropping attacks, similarly to (\ref{kr}).
\section{CV QKD over satellite channels}

The main challenge in the satellite-based communication, and particularly the quantum one, is the extremely strong attenuation levels, which are much higher than the typical loss in the terrestrial fiber and free-space links in which QKD was mostly tested. 

The total level of signal loss in a satellite link can widely vary and depends on the type of satellite and technical specifications of the channel realization. Indeed, in the recent experiment with measurement of the quantum-limited signal from a geostationary satellite using the homodyne detection, the total loss of 69 dB was observed \cite{Guenthner2017} with an aperture of 27 cm. The loss can be reduced to 55 dB once a bigger aperture of 1.5 m is used. Alternatively, the channel loss from a low Earth orbit (LEO) satellite can be substantially smaller and as low as 31 dB for an Alphasat-like satellite at a distance of 500 km. The loss can be reduced to about 20 dB by using larger receiving apertures \cite{Guenthner2017}. Therefore it is important to assess applicability of CV QKD in various scenarios, mainly resulting in different optical link attenuation levels.

\subsection{Quantum channel and protocol parameters}
The protocols are essentially influenced by the excess noise on the channel \textit{output}, $\epsilon$, further fixed to $10^{-4}$ shot-noise units (SNU), which are the vacuum quadrature fluctuations. This complies with the experiment in the 100-km optical fiber with the total attenuation of -20 dB, where the noise on the channel \textit{input} was estimated as $3\%$ SNU \cite{Huang2016} and with the recent experiment in the 300 km long low-loss fiber with the total attenuation of -32 dB, where the noise at the channel \textit{input} was estimated in the worst case as $3.8\%$ SNU \cite{zhanglong}. Note that the excess noise is mainly concerned with imperfect parameter estimation from the homodyne data on the receiving side of the protocol (even if the channel noise is physically absent, the pessimistic assumption on the level of noise, related to noise estimation error, results in effectively non-zero level of channel noise in order to comply with the required probability of failure of channel estimation procedures \cite{Ruppert2014}), which substantially depends on the stability of the set-up. It is essential that we fix the noise at the channel output contrary to the standard approach in CV QKD, when noise was fixed as relates to the channel input and then scaled by the channel attenuation, making the assessment of the channel noise in long-distance channels too optimistic. Alternatively and taking into account the fact that the excess noise atop of the calibrated electronic noise of the detector appears most likely due to the imperfect estimation of a noiseless quantum channel, one may assume passive eavesdropping such that no untrusted channel excess noise is present. In the case of satellite-based links, where line of sight between the sender and the receiver suggests the absence of equipment capable of active eavesdropping, this is a particularly sensible assumption and it was applied recently for feasibility study of DV QKD over satellite links \cite{Vergoossen2019}. We also take this assumption into account in CV QKD by assuming the excess noise to be trusted (\textit{i.e.}, being out of control by an eavesdropper) and including it in the state purification using the scheme similar to the entangling cloner with a strongly unbalanced coupling to the signal prior to detection \cite{Usenko2016}.

For the data ensemble size, we rely on the typical passage time of 300 seconds observed for the Micius quantum satellite \cite{Liao2018}. Assuming repetition rate of a CV QKD system to be of order of GHz, which is challenging but feasible with the current technology \cite{Zhang2018}, we may expect $N=10^{11}$ data points acquired during a satellite passage, half of which will then contribute to the key. Lastly, post-processing efficiency is taken $\beta=0.95$, complying with the currently available algorithms \cite{milicevic2018quasi}. 

The maximum tolerable channel attenuation gives the idea of the protocols applicability independently of the aperture setting and random atmospheric disturbances, \textit{i.e.} based only on the link optical budget. The results are summarized in Table \ref{restab} for given security assumptions and protocol parameters (signal states and clock rates). 
\begin{table}
\begin{tabular}{r|ccc|ccc}
													& \multicolumn{3}{c|}{Coherent states} & \multicolumn{3}{c}{Squeezed states} \\
													& 100 MHz     & GHz       & 10 GHz     & 100 MHz     & GHz       & 10 GHz    \\ \hline
    Active collective attack  & 23 dB       & 24 dB     & 25 dB      & 29 dB       & 30 dB     & 31 dB     \\
Passive collective attack & 27 dB       & 32 dB     & 37 dB      & 33 dB       & 37 dB     & 42 dB     \\
Active individual attack      & 29 dB       & 34 dB     & 39 dB      & 33 dB       & 37 dB     & 43 dB     \\
Passive individual attack     & 29 dB       & 34 dB     & 39 dB      & 33 dB       & 38 dB     & 43 dB   

\end{tabular}
\caption{Tolerable levels of channel attenuation (rounded) for various security assumptions and optimized CV QKD protocol settings. Excess noise on the channel output is fixed to $\epsilon=10^{-4}$ SNU, and optimal squeezing is $V_S\geq 0.1$ SNU. During a passive attack the excess noise is presumed to be trusted.}
\label{restab}
\end{table}

Evidently, in the assumption of passive eavesdropping there is no substantial difference between collective and individual attacks. On the other hand, when the channel excess noise is assumed to be untrusted, relaxing the assumptions on the possible attacks from collective to individual ones can substantially extend the tolerable loss. Note that the use of squeezed states typically increases the tolerable channel attenuation by 4-6 dB depending on the attack assumption (the better improvement being observed upon more strict collective attacks). Furthermore, with the high repetition rate the squeezed-state protocol can tolerate from 30 to 44 dB of channel attenuation depending on the attack assumption, making it potentially feasible in the geostationary scenario.

Performance of CV QKD over short-range (terrestrial) free-space links can be essentially limited by quantum channel transmittance fluctuations \cite{Usenko2012,pirandola2020limits}, also referred to as fading, which are mainly caused by the atmospheric turbulence effects of beam wander \cite{Vasylyev2012}, when a beam spot travels around the receiving aperture. The channel fading then results in increase of the channel noise, detected on the receiver \cite{Usenko2012}. Even though this effect will be partially compensated for in the long-distance realization of CV QKD, where beam spot drastically expands during the propagation, which results in channel stabilization at the cost of increasing the overall loss \cite{Usenko2018} as well as by active beam tracking and stabilization systems, it must be taken into account in the realistic CV QKD security analysis. However, as the residual transmittance fluctuations due to atmospheric effects are slow (of the order of KHz \cite{Ursin2007}) compared to high achievable repetition rates of CV QKD systems (enabled by GHz rates of homodyne detectors \cite{Zhang2018}), the fading can be further compensated for by properly grouping the data according to estimated relatively stable transmittance values \cite{Ruppert2019}, which we also apply in our study. 

\subsection{Satellite-to-ground channel model}
\begin{figure}
    \centering
    \includegraphics[width=.5\linewidth]{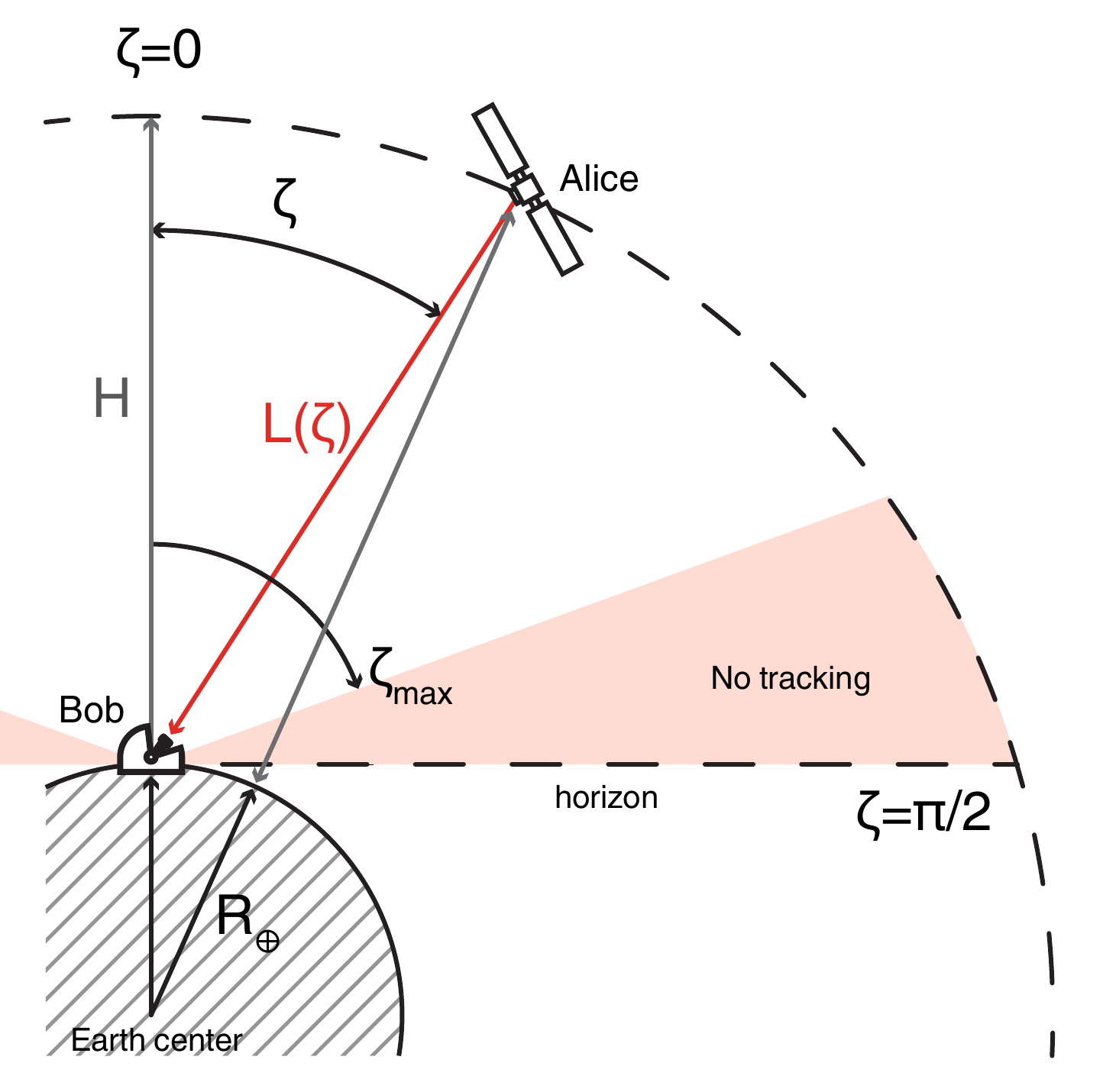}
    \caption{Geometrical representation of satellite-to-ground communication scheme. $H$ is the circular orbit altitude; $R_\oplus$ is Earth radius; $\zeta\in[-Pi/2,Pi/2]$ is the angle between the ray in the meridian plane, pointing above the observer (Bob), and the direct line connecting Bob and satellite (Alice), in practice limited to $\pm\zeta_{max}$; and $L(\zeta)$ is the slant range. }
    \label{geometry}
\end{figure}

In our study we consider downlink satellite channels, as the turbulence effects, being destructive for CV QKD \cite{Usenko2012}, are less pronounced in this regime compared to the uplink scenario, where the signal is strongly affected by the atmosphere already in the beginning of the channel \cite{bourgoin2013comprehensive}. Optical Gaussian beam in the satellite-to-ground link is influenced by analytically predictable systematic and statistical effects occurring during the communication window. Aside from losses within the receiver optical system $\eta_{det}$, due to coupling and detection inefficiencies, systematic effects also include diffraction and refraction induced losses. The main source of loss is beam-spot broadening caused by diffraction. The broadening limits the maximal transmission efficiency $\eta_0$ as finite receiving aperture of size $a$ truncates and measures only a part of the incoming collimated beam with spot-size $W$. The latter is largely determined as $W=\theta_{d}L(\zeta)$ by divergence of the beam $\theta_{d}$ and the channel length $L(\zeta)$. Line-of-sight distance, also referred to as slant range, between the observer and the satellite depends on the exact position of the latter. In following we assume a perfectly circular Low-Earth orbit within observers meridian plane, characterized by the altitude above the ground $H$, ranging from 200 km to 2000 km. The slant range is obtained as 

\begin{equation}
    L(\zeta)=\sqrt{H^2+2HR_\oplus+R_\oplus^2\cos^2\zeta}-R_\oplus\cos\zeta,
\end{equation}

where $R_\oplus$ is Earth radius and $\zeta$ is zenith angle \textit{i.e.} between the line pointing in opposite direction to the gravity from the observer and the slant range, see also Figure \ref{geometry}. Additionally, air density and optical refractive index change with the altitude causing the bending of the beam and making the overall traveling distance longer than the straight geometrical slant range. The refraction induced elongation also depends on the signal wavelength, geographical position and altitude of the receiver, as well as atmospheric conditions (relative humidity, temperature, wind, pressure). We assume the communication window is established up to the zenith angle $\zeta_{max}=70^\circ$, when the difference between actual and perceived slant ranges is small \cite{vasylyev2019satellite}. While such limitation reduces the communication window and the size of data block, it also avoids contributions from the longest propagation distance overall and through the thickest air mass \cite{Bedington2017, lee2019updated}. 
Duration of communication window, \textit{i.e.} total time in view of the satellite $t$, determines the amount of accumulated data points $N$ for given source repetition rate. The time in view is calculated from geometrical considerations and satellite orbital velocity, which for circular orbit in the observers meridian plane can be simplified as \cite{larson1999space, lissauer2013fundamental}:

\begin{equation}
    t\approx 2\frac{(R_\oplus+H)^{3/2}}{\sqrt{G\cdot M_\oplus}}\left(\zeta_{max}-\arcsin{\left[\frac{R_\oplus}{R_\oplus+H}\sin{\zeta_{max}}\right]}\right),
\end{equation}
where $G$ is the gravitational constant, and $M_\oplus$ is Earth mass.

The volume of air mass is related to the atmospheric extinction ratio $\eta_{ext}$, which is a measure of Rayleigh scattering, scattering due to aerosols, and molecular absorption experienced by the signal beam in terrestrial atmosphere \cite{palmer2001ccd}. The value of extinction ratio $\eta_{ext,\zeta=0}$ depends on the the signal wavelength, air constituents, temperature and weather conditions, and its change with increase of the path length through the atmosphere (in terms of zenith angle) can be approximated (for angles up to $\zeta_{max}$ where refraction effects are small) as  \cite{hardie1962astronomical, tomasi2014calculations}:

\begin{equation}
    \eta_{ext}(\zeta)=\eta_{ext,\zeta=0}^{sec(\zeta)}.
\end{equation}

The ratio at zenith can be obtained based on MODTRAN atmospheric transmittance and radiance model \cite{berk2014modtran}, which for rural or urban sea-level Mid-latitude location with clear sky visibility yields $\eta_{ext}=0.908$. 

Aside from aforementioned effects, refraction and diffraction are also caused by wind shear and temperature fluctuations and consequently by spatial and temporal variations of refractive index in the channel. Such perturbations result in scintillation, deviation of beam-spot from the center of the receiving aperture, and deformation of the beam-spot. For satellite links, the beam-spot radius is always significantly larger than the aperture, \textit{i.e.}, $W>a$, which allows us to ignore the deformation of the Gaussian beam profile, hence making beam wandering the dominant effect governing the fluctuation statistic of the channel transmittance.

    \begin{figure}
        \centering
        \includegraphics[width=.75\linewidth]{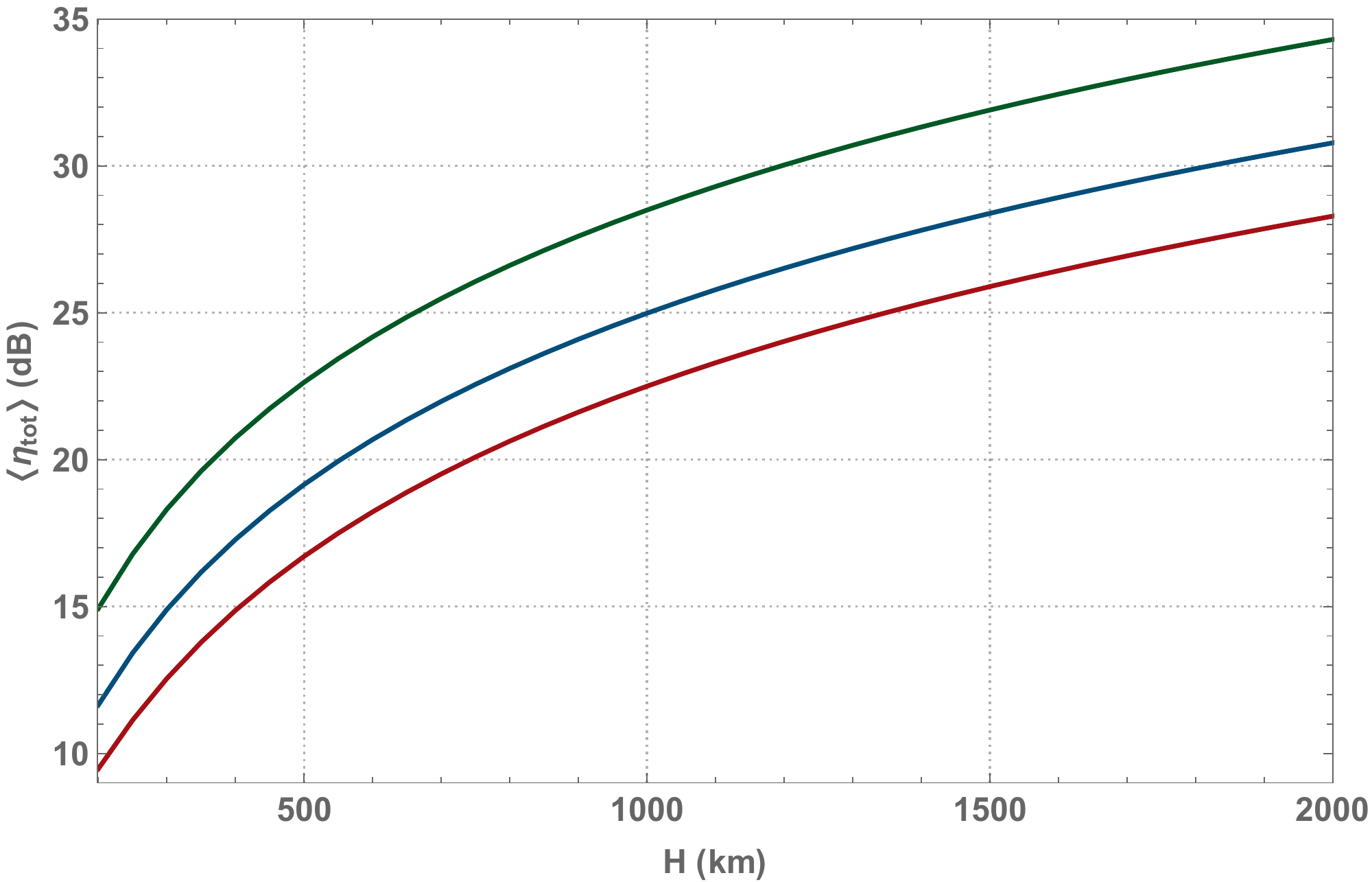}
        \caption{Mean signal loss in satellite-to-ground channel dependence on orbit altitude $H$ with receiver aperture radius $a=0.5,\,0.75,\,1\,m$ (from top to bottom respectively). Single pass transmittance distribution is assembled from simulated individual links at every zenith angle within $-\zeta_{max}<\zeta<\zeta_{max}$. }
        \label{fig:etaperpass}
    \end{figure}%

The maximal transmission efficiency is reached when incoming signal is perfectly aligned with the receiving aperture ($r=0$), and is defined by the ratio $a/W$ of the aperture and beam spot sizes as follows

    \begin{equation}
        \eta_{0}=1-\exp\left[-2\left(\frac{a}{W}\right)^2\right].
    \end{equation}

Efficiency $\eta\in[0,\eta_0]$ decreases with the increase of deflection distance $r$, with approximate analytical solution \cite{Vasylyev2012} being

    \begin{equation}
        \eta=\eta_0\exp\left[-\left(\frac{r}{R}\right)^\lambda\right],
    \end{equation}

where $\lambda$ and $R$ are shape and scale parameters respectively: 

   \begin{equation}
        \lambda = 8\left(\frac{a}{W}\right)^2\frac{\exp\left[-4\left(\frac{a}{W}\right)^2\right]I_1\left[4\left(\frac{a}{W}\right)^2\right]}{1-\exp\left[-4\left(\frac{a}{W}\right)^2\right]I_0\left[4\left(\frac{a}{W}\right)^2\right]}     \bigg\{\ln\left[\frac{2\eta_0}{1-\exp\left[\-4\left(\frac{a}{W}\right)^2\right]I_0\left[4\left(\frac{a}{W}\right)^2\right]}\right]\bigg\}^{-1},
    \end{equation}

    \begin{equation}
        R=a\Bigg\{\ln\left[\frac{2\eta_0}{1-\exp\left[-4\left(\frac{a}{W}\right)^2\right]I_0\left[4\left(\frac{a}{W}\right)^2\right]}\right]\Bigg\}^{-1/\lambda},
    \end{equation}

where $I_n$ is $n$-th order Bessel function. The position of deflected beam center is assumed to be normally fluctuating around the aperture center \cite{Vasylyev2018}, and described by random transverse vector $\textbf{r}_0$:

    \begin{equation}
        \mathcal{P}(\textbf{r}_0)=\frac{1}{2\pi \sigma^2_{BW}}     \exp\left[-\frac{\textbf{r}_0^2}{2\sigma^2_{BW}}\right],
    \end{equation}

where beam-wandering variance is limited by tracking accuracy and beam stabilization $\sigma_{BW}=\theta_p L(\zeta)$, and $r=|\textbf{r}_0|^2$. The analytical form of the probability distribution of atmospheric transmittance is the log-negative Weibull distribution \cite{Vasylyev2018}:

    \begin{equation}
        \mathcal{P}(\eta)=  \frac{R^2}{\lambda\eta\sigma^2_{BW}} \left(\ln\frac{\eta_0}{\eta}\right)^{2/\lambda-1} \times \nonumber \exp\left[-\frac{R^2}{2\sigma^2_{BW}}\left(\ln\frac{\eta_0}{\eta}\right)^{2/\lambda}\right].
    \end{equation}

We simulate the values of $\eta$ for each zenith angle $\zeta\in[-\zeta_{max},\,\zeta_{max}]$ within the communication window with respect to the clock rate of the CV QKD protocol, and combine it with systematic receiver loss $\eta_{det}$, and respective extinction ratio $\eta_{ext}\,(\zeta)$ to obtain an overall mean transmittance $\langle\eta_{tot}\rangle$ (as well as $\langle\sqrt{\eta_{tot}}\rangle$) of the satellite pass. Hence the transmittance statistics during an overall communication window consists of contributions from  individual simulated free-space channels at every permitted zenith angle $\zeta$. Both $\langle\eta_{tot}\rangle$ and $\langle\sqrt{\eta_{tot}}\rangle$ govern the evolution of a covariance matrix of the state shared between Alice and Bob over the fluctuating channel \cite{dong2010continuous}. Note that we evaluate the mutual information between the trusted parties, $I_{AB}$, from the overall covariance matrix, averaged over the whole transmittance distribution (similarly to evaluating the Holevo bound), which we observe to be lower than the average mutual information, hence being the pessimistic estimate. Following is the list of parameters used in the simulation: the receiver aperture radius $a=0.5\,(m)$ (respective plots shown in green color) $0.75\,(m)$ (shown in blue), and $1\,(m)$ (red), wavelength $\lambda=1550(nm)$, detection efficiency $\eta_{det}= -3\,dB$ \cite{Bedington2017}, tracking and pointing accuracy $\theta_p=1.2\,(\mu rad)$ \cite{liao2017satellite}, and beam divergence $\theta_d=10\,(\mu rad)$ \cite{Liao2018}. The resulting mean transmittance for a single pass at altitude $H$ is shown in Figure \ref{fig:etaperpass}. 

\subsection{Security evaluation}
    
\begin{figure}[t]
\begin{tabular}{ll}
\includegraphics[width=0.49\linewidth]{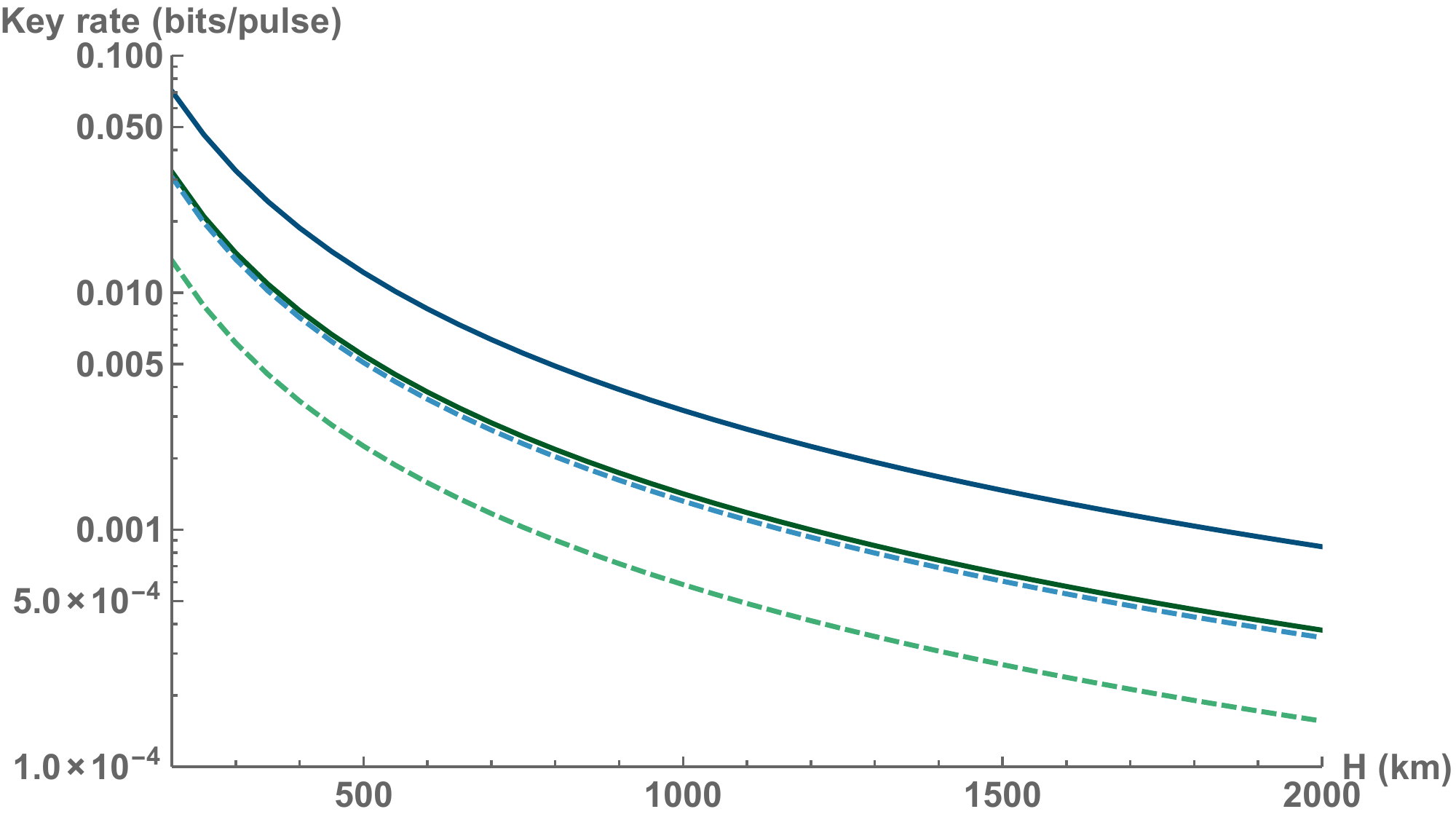}	
\includegraphics[width=0.49\linewidth]{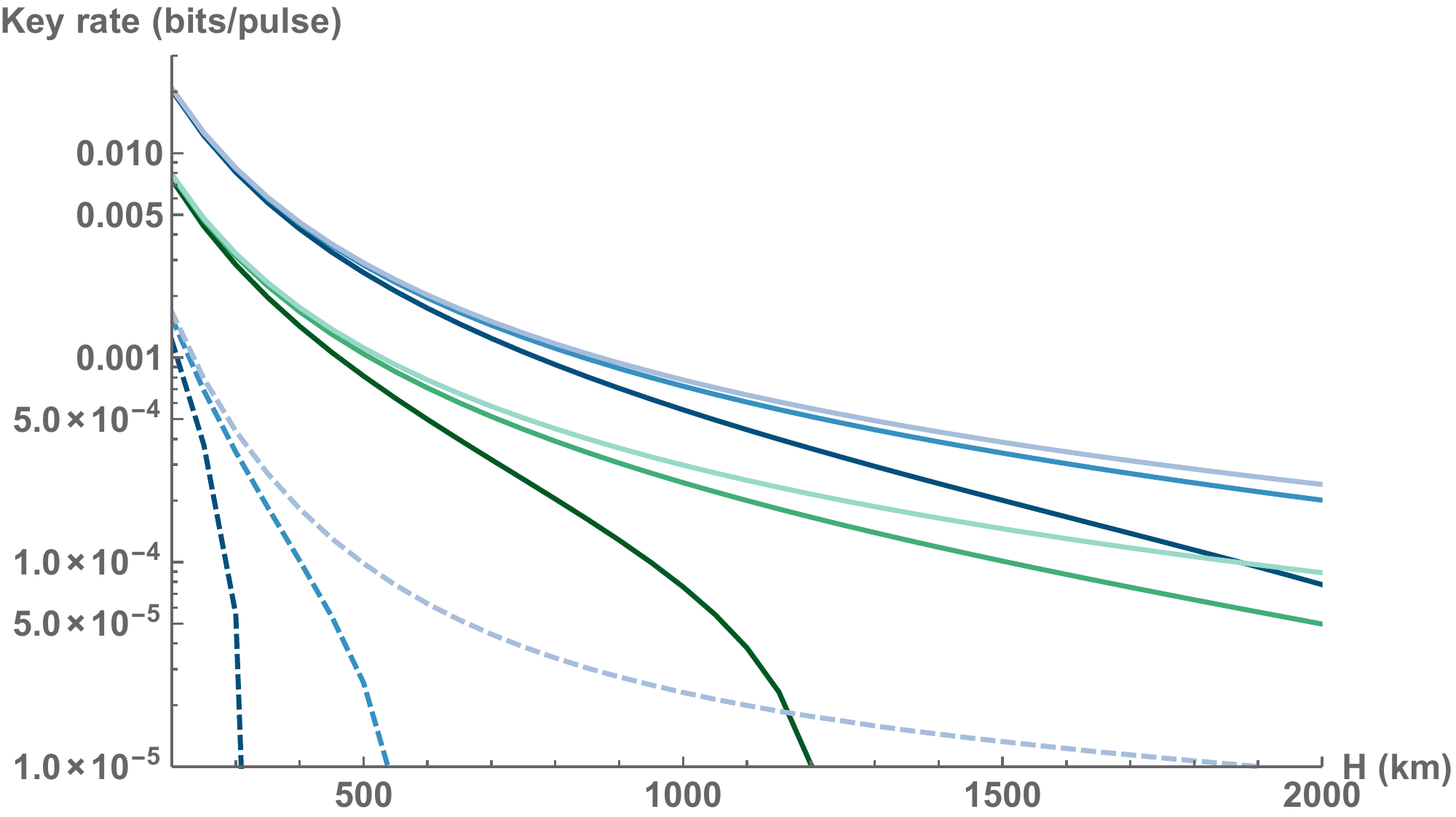}
\end{tabular}
\caption{(Left) Asymptotic lower bound on the key rate (in bits per channel use) of the optimized coherent-state (dashed) and squeezed-state CV QKD protocols (solid) secure against active individual attacks for a single pass of a LEO satellite with orbit altitude $H$ and receiver aperture radius $a=0.5,\,0.75m$ (green and blue lines respectively). Channel excess noise at the output $\epsilon=10^{-4}$ SNU. (Right) The key rate (in bits per channel use) of optimized coherent-state (dashed) and squeezed-state protocol (solid) under passive collective attacks for a single pass of a LEO satellite with orbit altitude $H$, and receiver aperture radius $a=0.5,\,0.75 m$ (green and blue lines respectively). Note that the coherent-state protocol can only be securely established with larger aperture $a=0.75m$. The overall block size depends on the length of communication window (given by the orbit altitude $h$) and the clock rate (from bottom to top) 100 MHz, 1 GHz, and 10 GHz. Trusted channel noise at the output is $10^{-4}$ SNU. Both squeezing and modulation variance are optimized (with optimal squeezing limited by a feasible value as $V_S\geq 0.1$) in order to establish a secure key regardless of the altitude.} 
\label{KR1}
\end{figure}

\begin{figure}[t]
    \centering
        \begin{tabular}{ll}
            \includegraphics[width=0.49\linewidth]{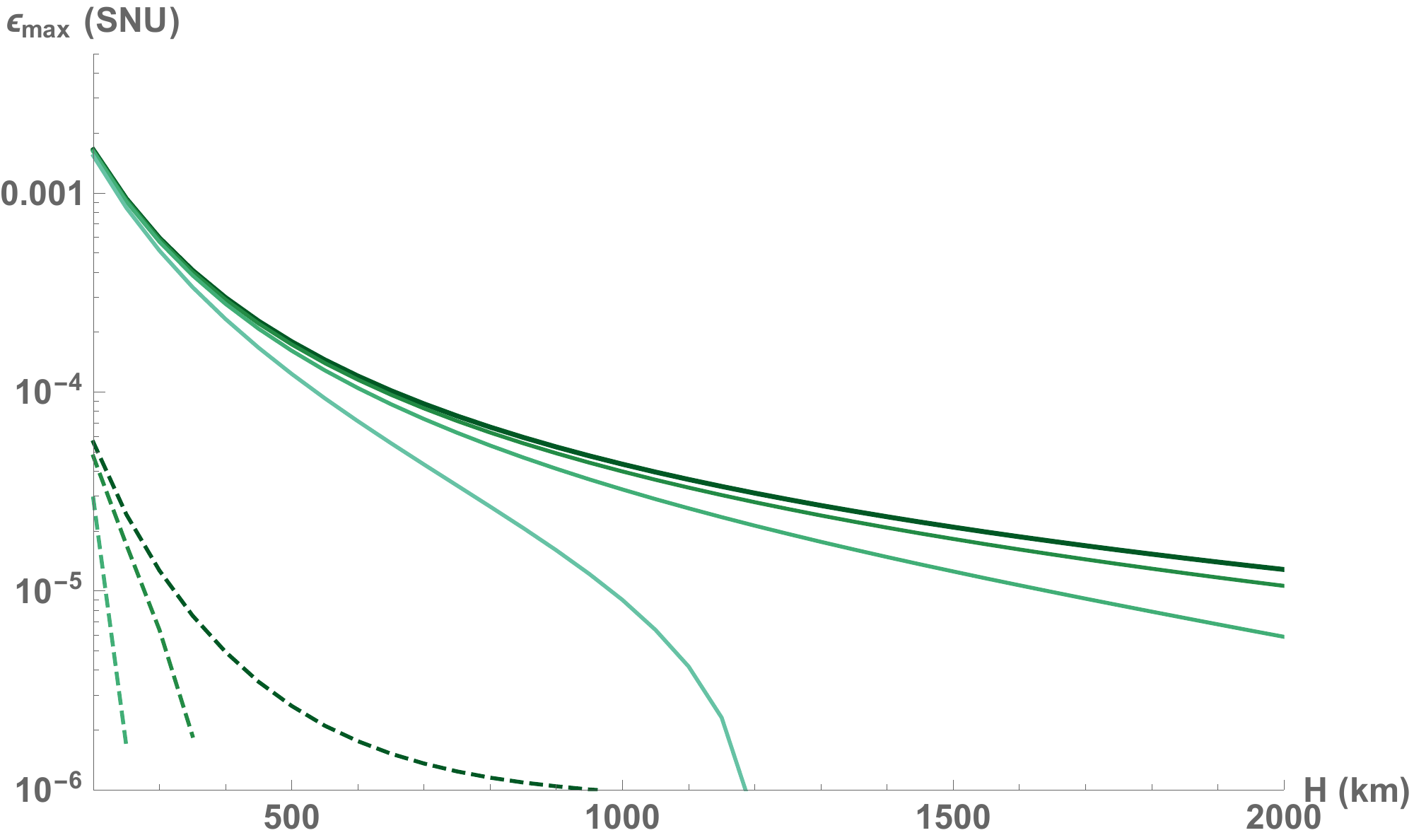}	
            \includegraphics[width=0.49\linewidth]{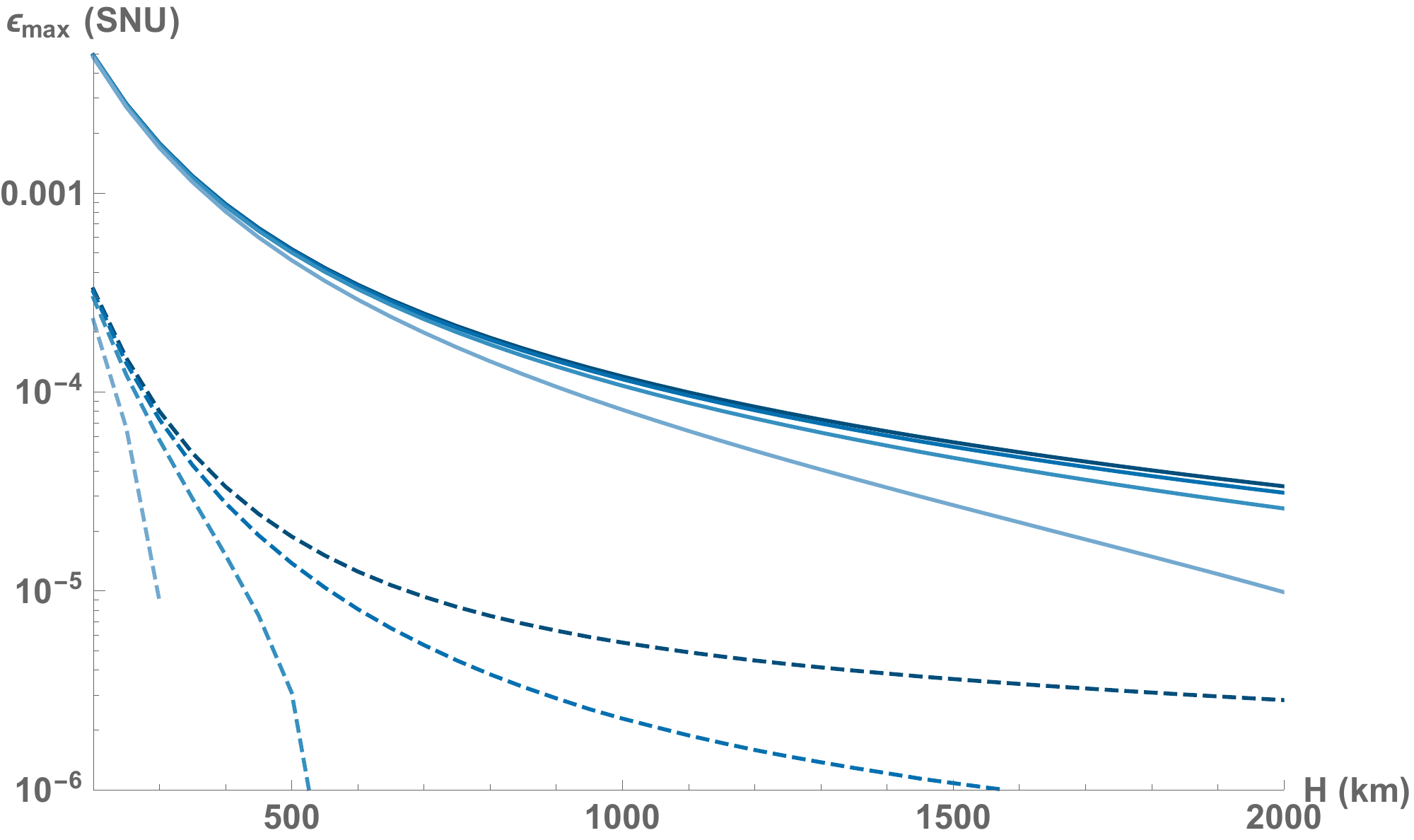}
        \end{tabular}
    \caption{Maximal tolerable excess noise in the case of active collective attacks on squeezed-  or coherent-state protocols (solid and dashed lines respectively) with receiver aperture size $a=0.5$ (left) and $0.75$ (right). From top to bottom: asymptotic regime, finite-size regime with repetition rate (from bottom to top) 100 MHz, 1 GHz, and 10 GHz. Squeezing $V_S\geq0.1$ and modulation variance $V_M$ are optimized.}
    \label{emax}
\end{figure}

\begin{figure}
\begin{tabular}{ll}
\includegraphics[width=0.48\linewidth]{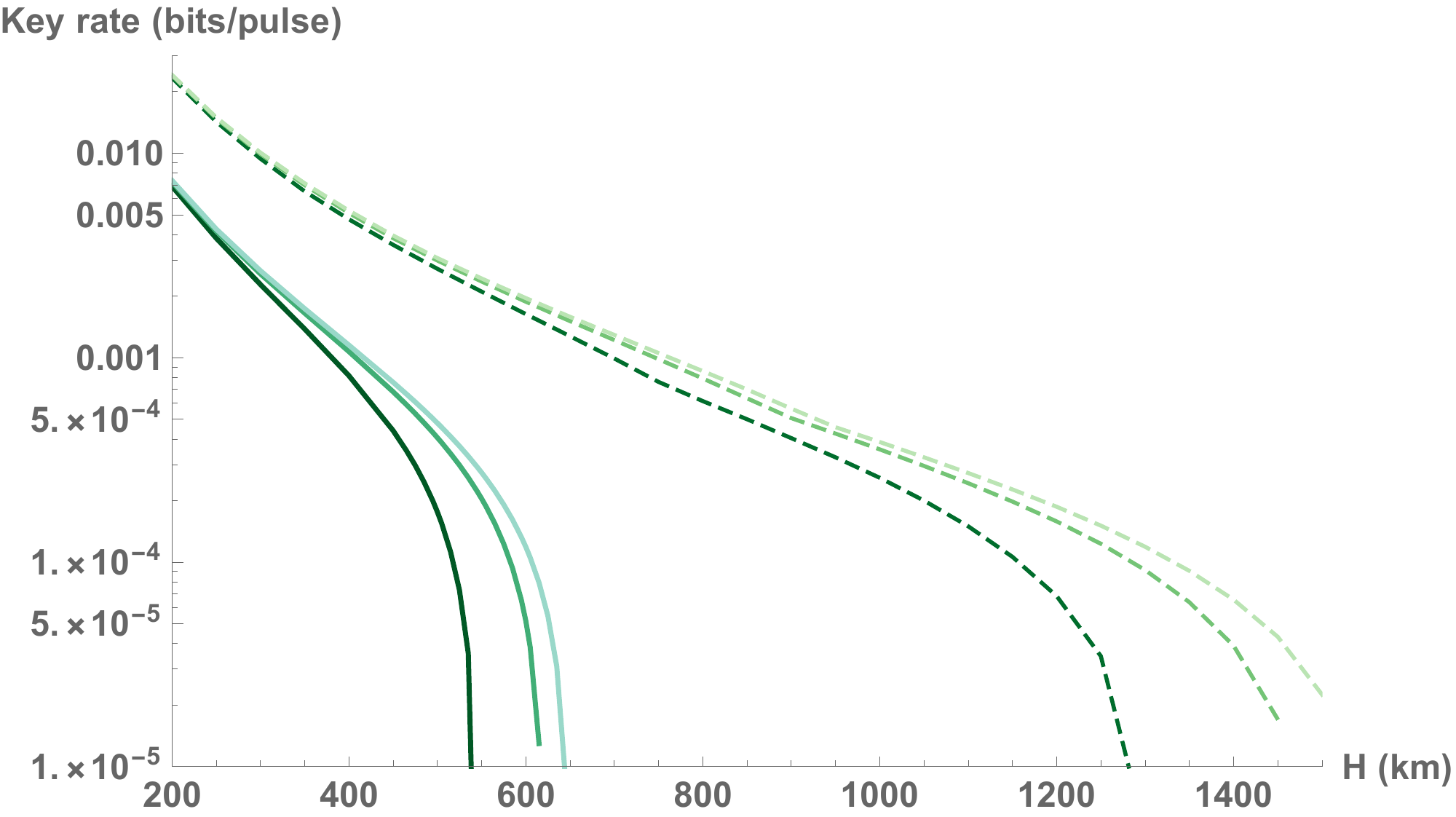}
\includegraphics[width=0.48\linewidth]{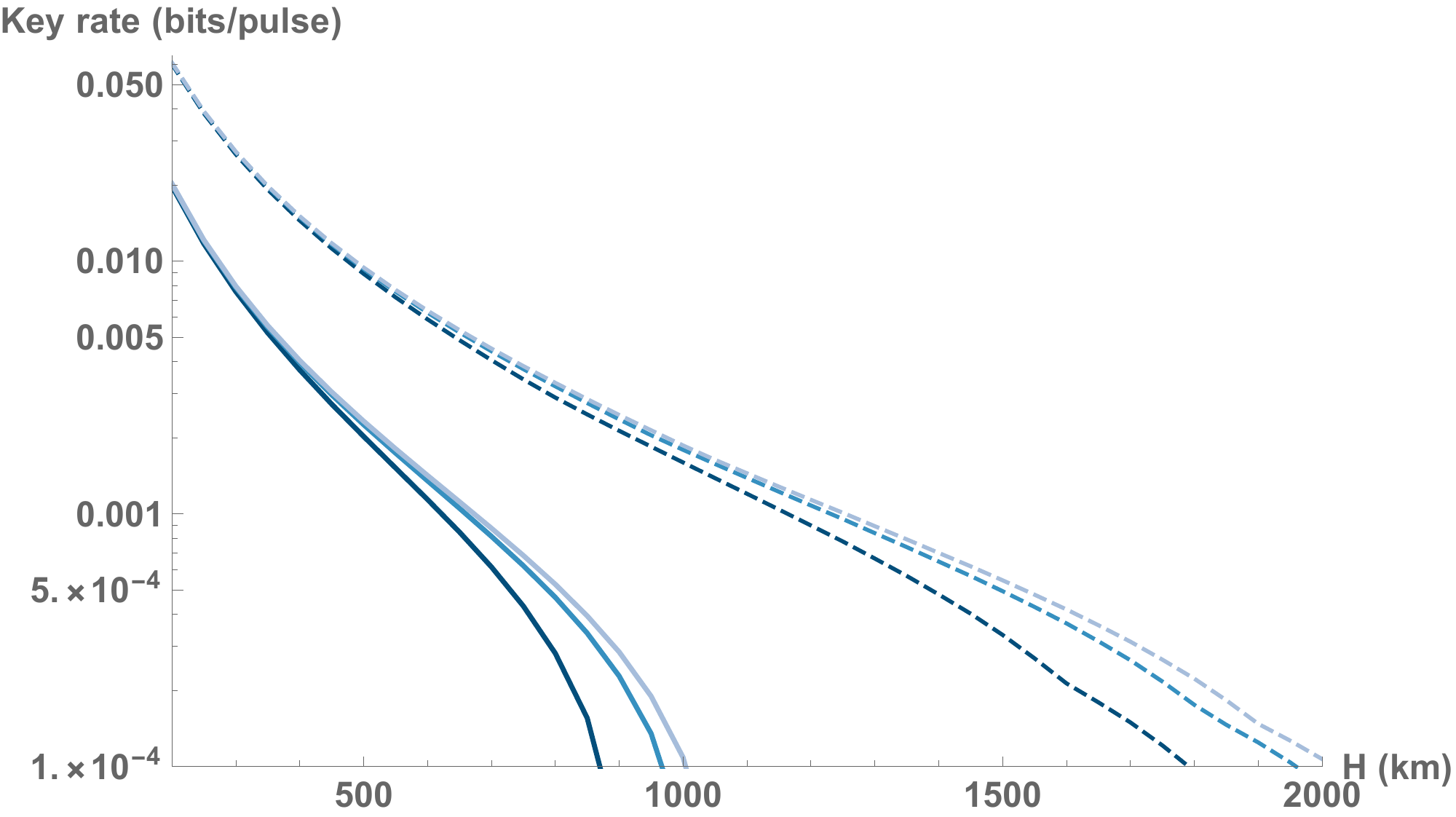}	
\end{tabular}
\caption{Secure key rate without channel subdivision (solid lines) and with channel subdivision in 3 segments (dashed lines) versus orbit altitude $H$ (km), secure against collective attacks in the finite-size regime with number of data points determined by (from darker to lighter shade) repetition rate $100 MHz,\, 1 GHz,\, 10 GHz$ for the squeezed-state protocol with optimized squeezing $V_S$$\geq 0.1$ in the presence of untrusted excess noise $\epsilon=10^{-4}$ related to the channel output. Aperture size $a=0.5m$ (left) and $a=0.75m$ (right). 
\label{clusters}}
\end{figure}

We first assess the security by looking at the impact of individual attacks in asymptotic regime (see Figure \ref{KR1} left). In this regime, security can be established for every LEO satellite altitude and feasible squeezing $V_S\geq0.1$ (SNU) provides a noticeable rate gain. Note that under the individual attacks, the higher levels of squeezing always translate into a higher secure key rate, which is not the case for collective attacks where the transmittance fluctuations limit the applicable values of squeezing \cite{Derkach2020} and require active squeezing optimization based on the estimated atmospheric transmittance distribution. The optimization of the latter is a crucial step required for the establishing of secure key rate based on the data generated from a single pass of the satellite. 

The performance of the optimized CV QKD protocol under passive collective attacks with trusted noise $\epsilon=10^{-4}$ is depicted in Figure \ref{KR1} (right). The impact of finite-size effects reduces with an increase of the altitude $H$ as the communication window gets longer and consequently the overall block size $N$ gets larger. Low altitude satellite downlinks exhibit less mean attenuation, but such advantage is partially offset by larger confidence intervals of estimated channel parameters and shorter raw key. While optimization of estimation block-size $N-n$ can lengthen the key to some extent, increasing the repetition rate of the system is necessary to greatly extend the raw key. Higher repetition rate is especially crucial for the coherent-state protocol that can be implemented at the altitudes above 500 km only with 10 GHz clock rates. Increasing the size of receiving aperture also leads to significant improvement of the secure key rate. 
\par 
Mean channel attenuation and fading noise, originating from transmittance fluctuations, diminish the tolerance to the channel excess noise, as shown in Figure \ref{emax}. Evidently, both protocols are extremely sensitive to excess noise and no secure key can be generated at any orbit altitude if the noise at the output of the squeezed-state protocol is $\epsilon\geq 2 \cdot10^{-3}$ SNU for $a=0.5$, or if the noise $\epsilon\geq 5 \cdot10^{-3}$ for $a=0.75$, with orders of magnitude lower values needed for security of the coherent-state protocol ($\epsilon\geq 6 \cdot10^{-5}$ at $a=0.5$, or $\epsilon\geq 4 \cdot10^{-4}$ at $a=0.75$). Note that with an increase of satellite orbit altitude the rate at which the protocol looses noise tolerance decreases. As the optical channel becomes longer it also becomes more stable \cite{Usenko2012, Usenko2018}, \textit{i.e.} both mean attenuation and fading (viewed as the variance of transmittance fluctuations $\langle\eta_{tot}\rangle-\langle\sqrt{\eta_{tot}}\rangle^2$) are simultaneously reduced. 

While the effect at LEO altitudes is more apparent for the coherent-state protocol, the same dependency can be expected for the squeezed-state protocol at higher altitudes (MEO or GEO). Furthermore, channel stabilization will be more pronounced with less accurate beam-tracking $\theta_{p}$ which directly limits the beam-wandering and consequently channel fading. 

Clearly, in order to operate under collective attacks with untrusted channel noise, the noise has to be limited to very low values. This can be achieved by proper control of the set-up or precise parameter estimation, however one can as well reduce the amount of fading noise by dividing the overall single-pass data block into a subset of smaller blocks \cite{Usenko2012}. Data clusterization with respect to channel attenuation allows to compensate the effect of channel transmittance fluctuations \cite{Ruppert2019}, however such post-processing can be demanding for satellite-based QKD and is not needed for slow systematic changes of transmittance during the satellite pass. In the current work we therefore adopt simpler albeit similar method by splitting the satellite tracked pass into a set of segments and generating the key for each one. This allows to achieve the following lower bound on the overall secure key rate as a weighed sum of the key rates from individual segments:

\begin{equation}
    K=\sum_i\text{max}\left\{0,\frac{n_i}{N}\left[\beta I_{AB}-\chi_{BE}-\delta(n_i)\right]\right\},
\end{equation}

where $n_i$ is the raw key length for a given segment $i$ with number of data points $N_i$, so that $\sum_iN_i=N$, $N_i-n_i$ of data points is used for segment channel estimation, and the weight is determined by the relative size of the segment $n_i/N$, with $N$ being the overall block size for a given satellite pass. The segments are chosen in accordance with the zenith angle $\zeta$, so that for $i=1,2,3$ we obtain 3 segments each containing measurement results at respectively $[-\zeta_{max},-2/3\,\zeta_{max})\cup(2/3\,\zeta_{max},\zeta_{max}]$,  $[-2/3\,\zeta_{max},-1/3\,\zeta_{max})\cup(1/3\,\zeta_{max},2/3\,\zeta_{max}]$, and $[-1/3\,\zeta_{max},1/3\,\zeta_{max}]$. The finite-size effects are stronger within each segment and transmittance fluctuations are substantial, yet reduced. Three segments are already
 sufficient to attain an enhanced positive secure key rate and extend the range of secure altitudes, as shown in Figure \ref{clusters}. For systems with smaller apertures this implies an effective increase of no tracking zone $\zeta_{max}$, as the segment characterized by the longer slant ranges $L(\zeta)$ might not contribute to the overall key. 

\section{Conclusions and discussion}
We studied applicability of CV QKD over satellite links considering coherent and squeezed-state protocols, taking into account realistic satellite passage, atmospheric effects, finite data ensemble size, system clock rate and data processing efficiency. We show that the protocols are very sensitive to channel noise at the respective loss levels so that either set-up stabilization resulting in drastic decrease of noise or relaxation of security assumptions to individual or passive collective attacks is required for implementation over the low Earth orbit satellites. Satellite pass segmentation provides another viable option to reduce  channel fading thus improving the secure key rate and allowing to establish the secure link with satellites at higher altitudes. We show that the use of squeezing can make the protocols more applicable in the satellite links, allowing for higher attenuation and levels of noise at the given security assumptions, and has to be optimized in the collective attacks scenario. On the other hand, in the case of trusted noise assumption and at high repetition rates, squeezed-state CV QKD can tolerate attenuation levels up to 42 dB, which may open possibility of use over geostationary satellites. The obtained results are promising for satellite-based QKD potentially applicable in daylight conditions.

\acknowledgments{
\textbf{Acknowledgments:} Authors acknowledge support from the project LTC17086 of INTER-EXCELLENCE program of the {Czech Ministry of Education}, project 19-23739S of the Czech Science Foundation and EU H2020 Quantum Flagship initiative project No. 820466, ~{CiViQ}.} 

\conflictofinterests{\textbf{Conflicts of Interest:} The authors declare no conflict of interest.} 



\bibliographystyle{mdpi}
\renewcommand\bibname{References}

\end{document}